\documentclass[12pt]{article}
\usepackage{subeqnarray}
\usepackage{epsfig}
\usepackage{amssymb}
\usepackage{hyperref}
\usepackage{graphicx}
\usepackage{amsmath}
\usepackage{amssymb}
\usepackage{amsfonts}
\usepackage{bbm} 

\newcommand{\eref}{Eq.~\ref}
\newcommand{\sref}{Sec.~\ref}
\newcommand{\tref}{Tab.~\ref}

\begin{document}


\thispagestyle{empty}
\begin{flushright}
 IFUP-TH/2008-33\\
October 2008 \\
\end{flushright}
\vspace{3mm}

\begin{center}
{\Large \bf A Generalized Construction for Lumps\\
and\\
Non-Abelian Vortices\footnote{Contribution to ``Continuous Advances in QCD 2008'', Minneapolis, May 2008. To appear in the Conference Proceedings.
}\\}
\vspace{15mm}

\normalsize { \bf {Walter~Vinci\footnote{walter.vinci(at)pi.infn.it}}}

\vskip 1.5em

{\small
 {\it Department of Physics ``E. Fermi'', University of Pisa}\\
and \\
{\it INFN, Sezione di Pisa,\\
  Largo Pontecorvo, 3, Ed. C, 56127 Pisa, Italy } }

\vspace{15mm}
%
%
{\bf Abstract}\\[5mm]
{\parbox{13cm}{\hspace{5mm}

We construct the general vortex solution in a fully-Higgsed, color-flavor locked vacuum of a non-Abelian gauge theory, where the gauge group is taken to be the product of an  arbitrary simple group and $U(1)$,  with a Fayet-Iliopoulos term. The strict correspondence between vortices and lumps in the associated NL$\sigma$M which arise in the limit of strong coupling is pointed out. The construction of the vortex moduli space is derived here as a consequence of this correspondence.



}}
\end{center}
\vfill
\newpage
\setcounter{page}{1} \setcounter{footnote}{0}
\renewcommand{\thefootnote}{\arabic{footnote}}

\section{Introduction}

Vortices play important roles in various areas of physics from  condensed matter physics 
to particle physics \cite{Achucarro:1999it} and cosmology  \cite{Vilenkin}. Most importantly, vortices could play a crucial role in the confinement mechanism of QCD \cite{'tHooft:1981ht}.

On the other hand, non-linear sigma models (NL$\sigma$M), and in particular the $\mathbb C P^1$ sigma model \cite{Polyakov:1975yp}, have been intensively studied because of their similarities with QCD, like asymptotic freedom, and non-perturbatively generated mass gaps.

Recently, there has been a significant progress in the understanding of non-Abelian vortices in the color-flavor locked vacuum of $U(N)$ gauge theories  \cite{Hanany:2003hp,Auzzi:2003fs}.
 Unlike Abelian vortices \cite{Abrikosov:1956sx},
they carry orientation moduli  in the internal space, in
addition to the usual position moduli. The most general 
Bogomol'nyi-Prasad-Sommerfield (BPS) vortex solutions and their moduli spaces have been
found  \cite{Eto:2005yh} and many other interesting features  have been extensively explored  \cite{Eto:2006db}. Most studies so far have  been restricted to the gauge group $U(N)$, with a few but notable  exceptions  \cite{Ferretti:2007rp,DKO} mainly devoted to the investigation of GNO duality\cite{Goddard:1976qe} from the vortex side.


This contribution is based on Ref. \cite{Eto:2008yi}, where we presented a simple framework to construct the most general non-Abelian BPS vortex solution in theories with an  arbitrary gauge group of type  $G=G' \times U(1)$. Here we will derive the same construction emphasizing the deep connections between vortices and lumps. 
 In fact, our prescription corresponds to a ``generalized'' rational map construction \cite{Polyakov:1975yp} that  can be used to construct both vortices and lump solutions in a wide class of gauge theories and non-linear sigma models.

We take  $G'$ to be a simple Lie group, but the method can be easily generalized to non-simple groups. The cases of classical groups $G'=SU,SO,USp$ will be worked out in more detail. 
\section{The Model and BPS Vortex Equations}

Let us start with presenting the class of models in which we will construct vortices. It is a non-Abelian gauge theory with gauge group $G=G'\times U(1)$, whose Lagrangian is given by
\begin{eqnarray}
\mathcal L &=&- \frac{1}{4e^2} F_{\mu\nu}^0 F^{0\mu\nu}
   - \frac{1}{4g^2} F_{\mu\nu}^a F^{a\mu\nu}
   + \left(\mathcal D_\mu H_A\right)^\dag \mathcal D^\mu H_A \nonumber\\
& & - \frac{e^2}{2} \left|H_A^\dagger t^0 H_A - \frac{v^2}{\sqrt{2 N}}\right|^2
   - \frac{g^2}{2} |H_A^\dagger t^a H_A|^2 \ ,
\label{bosonic} \end{eqnarray}
where
${\cal D}_{\mu} H = (\partial_{\mu} + i W_{\mu})H$,
$e$ and $g$ are the gauge coupling constants
for $U(1)$ and $G'$, respectively, and $A$ is the flavor index.  Apart from the gauge bosons  $W_{\mu} = W_{\mu}^0 t^0 + W_{\mu}^a t^a$  the matter
content of the model consists of  $N$
flavors of  Higgs scalar fields in the fundamental representation, 
 with a common $U(1)$ charge, written as a color-flavor mixed
$N \times N$ matrix $H$.  $t^0$ and $t^a$ denote the generators of $U(1)$ and $G'$ and are 
normalized as
\begin{eqnarray}
 t^0 = \frac{\mathbf 1_{N}}{\sqrt{2N}} \quad
  \text{Tr}( t^a t^b ) = \frac{\delta^{ab}}{2} \ .
\end{eqnarray}

The flavor symmetry of the model is $SU(N)_{\rm F}$. Though our discussion concerns mainly the bosonic system (\ref{bosonic}), the model is really to be considered as the (truncated) bosonic sector of the corresponding ${\cal N}=2$ supersymmetric gauge theory, which explains the particular form of the potential,  ensuring at the same time  its stability against radiative corrections. Another consequence of this is that we have several useful descriptions of the moduli space of vacua, which will be important in the upcoming discussion \cite{Luty:1995sd}:
\begin{subeqnarray}
\mathcal M_{G} & = & \left\{H \, | \, \text{D-term conditions}\right\}/G  \slabel{vac1}\\
                   & = & \left\{H\right\}//G^{\mathbb{C}} \slabel{vac2}\\
                   & = & \left\{I_G^{ij}, \,  \text{holomorphic} \, G\text{-invariants}\right\}/\{\text{algebraic relations}\slabel{vac3}\}
\end{subeqnarray}

Furthermore, $\mathcal N=2$ supersymmetry  implies the existence of BPS saturated vortex solutions which are supported by the non-trivial first homotopy group $\pi_1(G=G'\times U(1))=\mathbb Z$. The standard Bogomol'nyi completion for static, $x^{3}$-independent, configurations
\begin{eqnarray}
T &=& \int d^2 x \bigg[ 
  \frac{1}{2e^2} \left|F_{12}^0
   - e^2 \left(H_A^\dagger t^0 H_A - \frac{v^2}{\sqrt{2 N}}\right) \right|^2 + \nonumber\\
   & &\phantom{ \int d^2 x \bigg[}
   + 4\left|\mathcal D_{\bar z} H\right|^2
   + \frac{1}{2g^2}  \left|F_{12}^a
   - g^2 H_A^\dagger t^a H_A \right|^2
   - \frac{v^2}{\sqrt{2 N}} F_{12}^0 \bigg] \notag \\
&\geq& ~ - \frac{v^2}{\sqrt{2 N}} \int d^2 x \, F^0_{12} \ , \label{bpscompl}
\end{eqnarray}
gives the BPS equations for vortices.

We will focus our attention on the classical Lie groups $SU(N),\, SO(2M)$ and $USp(2M)$.
For  $G'= SO(2M),\, USp(2M)$ their group elements are embedded into $SU(N)$ ($N=2M$) by constraints of the form,   $U^T
J U = J$, where $J$ is the rank-2 invariant tensor
\begin{eqnarray}
 J = \left(\begin{array}{cc}
            \mathbf{0}_M & \mathbf{1}_{M} \\
 \epsilon \mathbf{1}_{M} & \mathbf{0}_M \end{array}\right)
,
\end{eqnarray}
where $\epsilon = +1$ for $SO(2M)$, while $\epsilon = - 1$ for $USp(2M)$. Specializing \ref{bpscompl} to these cases we get the  BPS vortex equations
\begin{subeqnarray}
 \mathcal D_{\bar z} H &= 0, \slabel{eq:BPSeq1}\\
 \frac{\sqrt{2 N}}{e^2}F_{12}^0 -  \left(\text{Tr}(HH^\dagger) - v^2\right)
 & = 0 \slabel{eq:BPSeq2},\\
 \frac{4}{g^2}F_{12}^a t^a - 
 \left(HH^\dagger- J^\dagger (HH^\dagger)^T J\right) &= 0,\slabel{eq:BPSeq3}
\label{eq:BPS}
\end{subeqnarray}
 where the complex coordinate $z \equiv x^1 + i x^2$ has been introduced. \eref{eq:BPSeq3}
reads for $G'=SU(N)$ instead: 
\begin{eqnarray} \frac{2}{g^2}F_{12}^at^a - \, \left[\, HH^\dag -
\frac{\mathbf{1}_N}{N} \text{Tr}(HH^\dag)\, \right] = 0 \ . \label{eq:BPSsun}\end{eqnarray}

\section{The Strong Coupling Limit and the BPS Lump Equations}

Another class of theories we are interested in, is the set of non-linear sigma models which can be obtained from the theories (3) by taking the strong coupling limit: $e,g \rightarrow \infty$. In a fully Higgsed vacuum, all gauge fields are massive and can be integrated out, and the limit is reliable. The resulting theory is an $\mathcal N=2$ $\text{NL}\sigma\text{M}$ in 3+1 dimensions whose target space is the vacuum manifold of the gauge theory: 
\begin{eqnarray}
 \mathcal L_{\text{NL}\sigma\text{M}}=g_{i,\bar j}\partial_\mu \phi_i \partial_\mu \bar \phi_j,& &  g_{i,\bar j}=\partial_{\phi_i}\partial_{\bar \phi_j} K(\phi_i,\bar \phi_j), \nonumber \\
\mathcal M_{\text{NL}\sigma\text{M}}&=&\mathcal M_{G}. \label{sigmadef}
\end{eqnarray}
The fields $\phi_i$ are a conveniently chosen set of independent coordinates on the target space while $K$ is the K\"ahler potential of the gauge theory.

This class of models also admits BPS, $x^3$-independent, stringy soliton called lumps. This kind of solitons are supported by a non-trivial homotopy group $\pi_2(\mathcal M_{\text{NL}\sigma\text{M}})$. It is indeed very easy to find the equations for lumps without going into the explicit computation of the K\"ahler potential. In fact, taking the strong coupling limit in (\ref{bosonic}) without explicitly intagrating out the gauge fields, we obtain a description of the models (\ref{sigmadef}) in terms of the fields of the original gauge theory $(H,W_{\mu})$. Lump equations are simply given by the strong coupling limit of the vortex equations (\ref{eq:BPS}).
\begin{subeqnarray}
\mathcal D_{\bar z} H &= 0, \slabel{eq:BPSlumpeq1}\\
\text{Tr}(HH^\dagger) - v^2
 & = 0, \slabel{eq:BPSlumpeq2}\\
 HH^\dagger- J^\dagger (HH^\dagger)^T J &= 0
  \slabel{eq:BPSlumpeq3}.\label{eq:BPSlump}
\end{subeqnarray}
Notice that  \eref{eq:BPSlumpeq2} and \eref{eq:BPSlumpeq3} are nothing but the D-term vacuum equations, whose solutions define $\mathcal M_{\text{NL}\sigma\text{M}}$.

A general property of the models (\ref{sigmadef}) is the classical scale invariance. Lumps develop a moduli space of degenerate solutions with arbitrary size and consequently there appear singular solutions with zero size, called small-lump singularities. Another way to understand these singularities is to notice that the NL$\sigma$Ms are well defined only on a fully Higgsed vacuum, while the theories (\ref{bosonic}) have a moduli space of vacua that usually includes many branches with different physical properties: Higgs and Coulomb branches, mixed Higgs-Coulomb phases in which the gauge symmetry is only partially broken. Lump solutions can be understood as maps from the $\mathbb C$-plane orthogonal to the $x^3$-axis, into the target space $\mathcal M_{\text{NL}\sigma\text{M}}$ and we expect singularities when this map hit a point which corresponds to a phase of the original gauge theory with an unbroken gauge symmetry. In fact, in this points, the appearance of massless particles gives rise to singularities in the K\"ahler potential of the gauge theory, and thus on the target manifold of the NL$\sigma$M. For example,  small-lump singularities are usually related to conical singularities in the target space\footnote{$\mathcal M_{\text{NL}\sigma\text{M}}$ can contain also different kind of singularities, like curvature singularities. They give rise to different kinds of singular lumps \cite{Eto:2008qw}.}.

Before giving the recipes for the construction of lumps, we will comment, in the next section, on the correspondence between vortices in the gauge theories (\ref{bosonic}) and lumps in the related NL$\sigma$M (\ref{sigmadef}). A consequence of this correspondence is that we get, as promised, the moduli space of vortices.
\section{The Vortex-Lump Correspondence}
\label{sec:correspondence}
The deep connection between vortices and lumps was first noticed by Hindmarsh \cite{Hindmarsh:1991jq}. In the extended Abelian Higgs model ``semilocal'' vortices \cite{Vachaspati:1991dz,Preskill:1992bf} appear, which are generalizations of the usual Abrikosov-Nielsen-Olesen (ANO) vortices: they possess zero modes related to an arbitrary size parameter. Semilocal vortices ``interpolate'' between ANO vortices and  $\mathbb CP^n$ lumps, in the sense that zero-size semilocal vortices are ANO (``local'') vortices, while semilocal vortices with increasing size become identical to lumps.

We consider this relationship from another, but equivalent, point of view  \cite{Shifman:2006kd}: taking the strong coupling limit $e,g \rightarrow \infty$ we map the gauge theory (\ref{bosonic}) into the sigma model (\ref{sigmadef}); at the same time, we  map any vortex appearing in the first theory into a lump of the latter:
\begin{eqnarray}
  G' \times U(1) \;  \text{Gauge theory} & \longrightarrow & \text{NL}\sigma\text{M} \; \text{on} \; \mathcal M_{G} \nonumber\\
   \text{Vortices} & \longrightarrow & \text{Lumps}.
   \end{eqnarray}
This correspondence can also be better understood if one consider more quantitative relations.

\textbf{BPS correspondence.} Vortices and lumps are BPS saturated objects. Their masses do not depend on the gauge couplings, being proportional to the central charge of the supersymmetry algebra: $T_{vor}=M_{lum}$

\textbf{Topological correspondence.} In a generic case in which the vacuum manifold is simply connected\footnote{This property of the vacuum manifold is the necessary condition for the existence of semilocal vortices  \cite{Preskill:1992bf}.}, the following relations hold: $\pi_2(\mathcal M_G)=\pi_2(\mathcal{M}_{vac \, man}/G)=\pi_1(G)$, which define a correspondence between the topological charges of vortices and lumps.

It is important, however, to notice that some vortex configurations will be mapped into singular lumps. For example, ANO vortices, with a typical size $\sim1/gv$, shrink to singular spikes of energy, and are mapped into small-lump singularities. Conversely, small-lump singularities are regularized into ANO vortices when finite gauge couplings are restored. The discussion above leads to the following relation among the moduli spaces of vortices and lumps  \cite{Shifman:2004dr}:
\begin{eqnarray}
\mathcal M_{vor}=\mathcal M_{lum}\oplus \{\text{singular lumps}\}
\end{eqnarray}


\section{The Generalized Rational Map Construction for Lumps}

After the general discussion of the previous Section, we are now ready to explicitly construct the moduli space of lumps arising from (\ref{sigmadef}).

We choose the fully Higgsed, color-flavor locked vacuum: $ H_{\rm vev}=\frac{v}{\sqrt{N}}{\bf 1}_N$. 
The $G' \times U(1) \times SU(N)_{\rm F}$ invariance of the theory is broken  to the global color-flavor
diagonal  $G'_{\rm C+F}$. The first step toward the solutions of \eref{eq:BPSlump} is to switch from the description (\ref{vac1}) for the moduli space of vacua to that of (\ref{vac2}). We can thus forget the D-term vacuum equations (\ref{eq:BPSlumpeq2}) and (\ref{eq:BPSlumpeq3}), at the prize of lifting the gauge symmetry $G$ to its complexification  $G^{\mathbb{C}}$.
\eref{eq:BPSlumpeq1} has now a very intuitive meaning. It can be thought as a ``covariant holomorphicity'' condition. This means that $H$ is holomorphic, up to a complexified gauge transformation:
\begin{eqnarray}
 H = S^{-1} H_0 (z) = S_e^{-1} {S'}^{-1}H_0 (z) \ \label{H0def},
\end{eqnarray}
where $S\in G^{\mathbb C}$, while  $H_0(z)$ is a matrix whose elements  are holomorphic in 
$z$. We denote $H_0(z)$ the moduli matrix
 \cite{Isozumi:2004vg}, as it encodes all moduli parameters\footnote{To reconstruct the solution in terms of the original Higgs fields $H$, one has still to solve \eref{eq:BPSlump} as algebraic equations for $S$.}. \eref{H0def} does not fix completely $G^{\mathbb{C}}$. $H_0(z)$ is in fact only defined up to holomorphic gauge transformations $V(z)$:
\begin{eqnarray}
  H_0 \sim V(z) H_0 = V_{e}\,  V^{\prime}(z) H_0 \ , \quad  
   V^{\prime} \in G'^{\mathbb{C}},\quad V_{e}\in \mathbb C^*\ .   \end{eqnarray}

The next step is to consider the holomorphic invariants $I_{G}^{(i,j)}(H)$ made of $H$, which are invariant under $G^{\mathbb C}$, with $(i,j)$ labeling them. This is equivalent to changing our description of $\mathcal M_{G}$ from \eref{vac2} to \eref{vac3}. The key point here is that, thanks to \eref{H0def}, these invariant are also holomorphic functions of the coordinate $z$:
\begin{eqnarray}
 I^{(i,j)}_{G}(H)=I_{G}^{(i,j)}\left(S^{-1} H_0(z)\right)
 = I^{(i,j)}_{G}(H_0)(z).
 \label{eq:G'inv}
\end{eqnarray}
We have thus identified a lump as a holomorphic map from the complex plane $\mathbb C$ to the set of holomorphic invariants $I_{G}^{(i,j)}$\footnote{All the points at the infinity of $z$ are in fact identified, so that the map is in fact defined on $\mathbb CP^1$.}. 

The last step is to impose the following boundary conditions to the holomorphic invariants:
\begin{eqnarray}
I^{(i,j)}_{G}(H)\Big|_{|z|\rightarrow \infty} =
 I^{(i,j)}_{\rm vev}
 \label{eq:boundary}\,.
\end{eqnarray}
It is convenient to translate the above conditions in terms of the holomorphic invariants of $G'$, $I^j_{G'}(H_0)$:
\begin{eqnarray}
 I_G^{(i,j)} (H_0) \equiv 
\frac{I^i_{G'}(H_0)}{I^j_{G'}(H_0)}
 \label{eq:G'inv/G'inv}\ ,
\end{eqnarray}
where we must take ratios of invariants with the same $U(1)$ charge: $n_i=n_j$\footnote{The $U(1)$ charge $n_i$ is simply defined as: $I^i_{G'}(\alpha H_0)=\alpha^{n_i}I^i_{G'}( H_0)$, for any complex number $\alpha$.}. \eref{eq:G'inv/G'inv} defines the generalized rational map construction for lumps. 

 It is easy to check that \eref{eq:boundary} and \eref{eq:G'inv/G'inv} together imply the following:
\begin{eqnarray}
 I^i_{G'}(H_0)\Big|_{|z|\rightarrow \infty}= I^i_{\rm vev} \, z^{\nu n_i}
\end{eqnarray}
 where $\nu$ is an integer common to all invariants. As $I^i_{G'}(H_0(z))$ are holomorphic, the above condition means 
that $I^i_{G'}(H_0(z))$ are {\it polynomials} in $z$.
Furthermore,  $\nu\, n_i$ must be a positive integer for all $i$: 
\begin{eqnarray}
\nu \, n_i \in {\mathbb Z}_{+}
\quad \rightarrow \quad \nu = {k}/{n_0} \ , \quad k\in {\mathbb Z}_{+}
\ ,
\end{eqnarray}
where  (GCD  $=$ the greatest common divisor) 
\begin{eqnarray}
 n_0\equiv {\rm GCD}\{n_i~|~I_{\rm vev}^i \neq 0 \} \ .
\end{eqnarray}
Note that a $U(1)$ gauge transformation $e^{2\pi i /n_0}$ leaves $I^i_{G'}(H)$ invariant:
\begin{eqnarray}
 I^i_{G'}(H')=e^{2\pi i n_i/n_0}I^i_{G'}(H)=I^i_{G'}(H):
\end{eqnarray}
the phase rotation
$e^{2\pi i /n_0}\in {\mathbb Z}_{n_0}$
changes no physics, and  the true gauge group is thus $G={U(1)\times G'}/{{\mathbb Z}_{n_0}}$, where  ${\mathbb Z}_{n_0}$ is the center of $G'$. A simple homotopy argument tells us  that $1/n_0$
is the $U(1)$ winding for the minimal, $k=1$, topologically non-trivial configuration. 
Finally,  for a given  $k$ the following important relation holds
\begin{eqnarray}
 I_{G'}^i(H_0)=
 I^i_{\rm vev} z^{k n_i/n_0}+{\cal O}(z^{k n_i/n_0-1}) \ ,
 \label{eq:H0const}
\end{eqnarray}
which implies nontrivial constraints on $H_0(z)$. The set of all inequivalent  $H_0(z)$ satisfying \eref{eq:H0const} identify a lump solution, and thus defines the full moduli space.

\section{Moduli Space of Vortices}
The lump construction explained in the last Section is only partially complete, because one has to identify all singular lump solutions, and this is not easy in general. This problem disappears when we lift the construction for lump to that for vortices in the related gauge theory, according to the discussion of \sref{sec:correspondence}. As explained there, singular lumps are regularized into vortices as we turn on finite gauge couplings, and \eref{eq:H0const} now give the full and complete description of the moduli space of non-Abelian vortices.

To reconstruct the original Higgs fields $H=S^{-1}H_0$ we now have to solve the BPS differential equations \eref{eq:BPSeq2} and \eref{eq:BPSeq3} (or \eref{eq:BPSsun}) in terms of $S$, while from \eref{eq:BPSeq1} we find the gauge fields:
\begin{eqnarray}
W_1 + i W_2 &=& -  2 i \, S^{-1} \bar \partial S.
\end{eqnarray}
The tension of the BPS vortices can be written as
\begin{eqnarray}
T = - \frac{v^2}{\sqrt{2 N}} \int d^2 x ~ F^0_{12} = 2 v^2  \int d^2 x
~ \bar \partial \partial \,  \log(S_e S_e^\dagger) \ .
\end{eqnarray}
The asymptotic behavior $S_e\sim |z|^{\nu}$ then determines the tension
\begin{eqnarray}
 T = 2 \pi v^2 \nu \ ,
\end{eqnarray}
 thus $\nu$ has to be identified with the $U(1)$ winding number of the vortex configuration. 

Let us now apply the general discussion above to concrete examples.
For   $G'=SU(N)$, with $N$ flavors,
there only exists one invariant
\begin{eqnarray}
 I_{SU}=\det(H) \ , \label{eq:SUinvariant}
\end{eqnarray}
with charge $N$.
Thus the minimal winding is equal to  $1/N$ and
the condition for $k$ vortices is given by:
\begin{eqnarray}
 A_{N-1}: \det H_0(z)=z^k+{\cal O}(z^{k-1}), \quad
 \nu = k/N \ .
\label{eq:SUconstraint}
\end{eqnarray}
For  $G'=SO(2 M),USp(2M)$,
there are  $ N (N \pm 1)/2$ invariants
\begin{eqnarray}
 (I_{SO,USp})^{r}{}_s = (H^{\rm T}J H)^r{}_s,
 \quad 1\le r\le s\le N 
 \label{eq:I-SOUSp} \ ,
\end{eqnarray}
in addition to (\ref{eq:SUinvariant}). The constraints are:
\begin{align}
{C_M,D_M}&: H_0^T(z) J H_0(z) = z^k J + {\cal O}(z^{k-1})\ ,
 \; \ \,\, \nu =k/2 \ . 
\end{align}
 Thus, vortices in the $SO(2M)$ and $Usp(2M)$ theories are  quantized in half integers of the $U(1)$ winding \cite{Ferretti:2007rp}.

Explicitly,  the minimal vortices
 in $SU(N)$ and  $SO(2M)$ or $USp(2M)$ theories are given respectively  by the moduli matrices:
\begin{eqnarray}
 H_0 =
 \left(\begin{array}{cc}
 z -  a & 0 \\
 {\bf b} & {\bf 1}_{N-1}
 \end{array}\right)
,\
 \left(\begin{array}{cc} z{\bf 1}_M - { \bf A} & {\bf C}_{S/A} \\
         {\bf B}_{A/S}             & {\bf 1}_M
 \end{array}\right)
 \label{eq:SO,k=1semiloc}.
\end{eqnarray}
The moduli parameters are all complex. For $SU(N)$, $a$ is just a number;  ${\bf b}$ is a column vector. For $SO(2M)$ or $USp(2M)$, the matrix ${\bf C}_{S/A}$ for instance  is symmetric or antisymmetric, respectively. And vice versa for ${\bf B} $.  Moduli matrices for $SO(2M+1)$ as well as those for $k=2$ vortices in $SU, SO, USp$ theories, 
will soon be given explicitly in Ref. \cite{Eto:2008qw,EFKGNOW}.

The index theorem gives the complex dimension of the
moduli space 
\begin{eqnarray} \dim_{\mathbb{C}}\left(\mathcal{M}_{G',k}\right) = \frac{k\, N^2}{n_0} \
. \label{dim_moduli_space_generic} \end{eqnarray}
 This was obtained  in Ref. \cite{Hanany:2003hp} for $SU(N)$; 
a proof in other cases will be reported elsewhere
\cite{Eto:2008qw,EFKGNOW}. In all cases studied  we have checked that the dimension of the moduli space inferred from the moduli matrices agrees with the one given in \eref{dim_moduli_space_generic}. 

Except for the $SU(N)$ case,  our model has a non-trivial Higgs branch (flat directions).
 The color-flavor locked vacuum   $H_{\rm vev} \propto {\bf 1}_N$
is just one of the possible (albeit the most symmetric) choices for the vacuum;  our discussion can readily be
generalized to a generic vacuum on the Higgs branch. This fact, however,  implies that our non-Abelian vortices have semilocal moduli even for $N_{f}=N$.  In contrast to the Abelian or $SU(N)$ cases, moreover, they exhibit new, interesting phenomena such as ``fractional'' vortices \cite{EFKGNOW}.  

The generalization to exceptional groups can be done if the  invariant tensors of each group are known. Indeed, this is the case, and \tref{table:excepgroups} lists them all.

\begin{table}[ht]
\begin{center}
\begin{tabular}{|c|c|c|c|c|c|c|c|c|}
\hline
$G'$ & $A_{N-1}$& $B_M$ & $C_M, D_M$ &
   $E_6$& $E_7$& $E_8$& $F_4$& $G_2$
 \\ \hline
$R$&
 $N$& $2M+1$& $2M$& $27$& $56$& $248$& $26$& $7$
  \\ \hline
\small{\rm rank\;inv}&
  $-$ & $2$ & $2$ & $3$ & $2,4$ & $2,3,8$ & $2,3$ & $2,3$
 \\ \hline
$n_0$&
  $N$&$1$&$2$&$3$&$2$&$1$&$1$&$1$ \\
\hline
\end{tabular}
\caption{  The dimension of the fundamental
  representation ($R$), the rank of the other invariants
  \cite{Cvitanovic:1976am} 
and  the
  minimal tension ${\cal\nu} = 1/n_0$ i.e.~the center ${\mathbb Z}_{n_0}$ of
  $G'$. 
  The determinant of the $R \times R$ matrix gives one invariant
  with charge, dim $R$.} \label{table:excepgroups} 
\end{center}
\end{table}

\section{Local (ANO-like) Vortices}

For various considerations,   we are interested in knowing  
which of the moduli parameters describe the local vortices,  the ANO-type vortices with  
 exponential tails. The moduli space of fundamental local vortices, for example, is completely generated by the global symmetries of the vacuum (see below) and it can eventually survive non-BPS deformations that preserve these symmetries. Furthermore, recently it was found that local vortices correspond to the subset of solutions with the maximum number of normalizable moduli \cite{Shifman:2006kd}.
Local vortices correspond, as we mentioned, to small lump singularities in the lump construction. These configurations are obtained when the rational map construction is degenerate, e.g., when all the invariants $I^i_{G'}$ share a common zero:
\begin{eqnarray}
 I_{G'}^i=    (z-z_0)^{n_{i}/n_{0}} \,   I_{G'}'^i.   \label{eq:H0locinsert} \end{eqnarray}
The rational map in  \eref{eq:G'inv/G'inv}, does not feel any of these common zeros, which can be interpreted as insertions of singular spikes of energy, or local vortices. Configurations with only local vortices are obtained imposing a complete degeneration of the rational map:
\begin{eqnarray}
 I_{G'}^i(H_{0, {\rm local}})= \left[ \prod_{\ell=1}^{k}   (z-z_{0 \ell})   \right]^{n_{i}/n_{0}} \,   I^i_{\rm vev}\;.   \label{eq:H0constloc} \end{eqnarray}
For  $G'=SO(2M), USp(2M)$
with $I_{SO,USp}$ of \eref{eq:I-SOUSp}
we find that the condition for vortices to be of local type is
\begin{eqnarray}
 H_{0, {\rm local}}^T(z) J H_{0, {\rm local}}(z) = \prod_{\ell =1}^k
 (z-z_{0 \ell})  \, J \ .
 \label{eq:local-cond}
\end{eqnarray}

Let us now discuss  a few concrete examples. The general solution for the  minimal  vortex
(\ref{eq:SO,k=1semiloc}) for  $G'=\{SU(N),SO(2M), USp(2M)\}$ is reduced
to a local vortex if we restrict it to be of the
 form:
\begin{eqnarray}
 H_{0, {\rm local}} =
 \left(\begin{array}{cc}
 z - a & 0 \\
 {\bf b} & {\bf 1}_{N-1}
 \end{array}\right),\
 \left(\begin{array}{cc} (z - a){\bf 1}_M  & 0  \\
         {\bf B}_{A/S}             & {\bf 1}_M
 \end{array}\right)
 \label{eq:SO,k=1}.
\end{eqnarray}
The vortex position is given by  $a$, while  ${\bf b}$ for $SU(N)$ and ${\bf B}_{A/S}$ for
$SO(2M)$ or $USp(2M)$ encode the Nambu-Goldstone modes associated with the breaking of the color-flavor symmetry by the vortex $G'_{\rm C+F} \to H_{G'}$. The moduli spaces are
direct products of a complex number and the Hermitian
symmetric spaces
\begin{eqnarray}
 {\cal M}_{G',k=1}^{\rm local}
 \simeq {\mathbb C} \times G'_{\rm C+F}/H_{G'} \ ,
\end{eqnarray}
$H_{SU(N)}=SU(N-1)\times U(1)$ while $H_{SO(2M),USp(2M)} = U(M)$. The results for $SU(N)$ and $SO(2M)$ are well-known
 \cite{Hanany:2003hp,Auzzi:2003fs,Ferretti:2007rp}. The matrices (\ref{eq:SO,k=1}) describe just one
patch of the moduli space. In order to define the manifold
globally we need a sufficient number of patches. 
 The number of patches is $N$ for $G'=SU(N)$ and $2^M$ for $G'=SO(2M),USp(2M)$. The transition
functions correspond to the $V$-equivalence relations \cite{Eto:2005yh}. 
 In the case of $G'=SO(2M)$, 
the patches are given by permutation of the $i$-th and the ($M+i$)-th columns in (\ref{eq:SO,k=1}). 
We find that no regular transition functions connect the odd and even
 permutations (patches), hence the moduli space consists of 
 two disconnected copies of $SO(2M)/U(M)$ \cite{Ferretti:2007rp}. The complex
 dimensions of the moduli spaces are $\dim_{\mathbb
C}{\cal M}_{SO(2M),k=1}^{\rm local}=\frac{1}{2} M(M-1) + 1$ and $\dim_{\mathbb
 C}{\cal M}_{USp(2M),k=1}^{\rm local}= \frac{1}{2} M(M+1) + 1$.


%
%
%
%

\section{Conclusion}
We have given all the necessary tools to construct vortex solutions in a wide class of non-Abelian gauge theories. In fact, our method may be potentially extended to any gauge theory with any matter content. The only requirement is $\mathcal N=2$ supersymmetry. The same method gives also a generalization of the rational map construction for lump, which can be applied to any NL$\sigma$M whose target space is the vacuum manifold of an $\mathcal N=2$ gauge theory.
Our method can also be extended to other BPS solitons  such as domain walls, monopoles and instantons, and hopefully opens powerful new windows for their investigation.

\section*{Acknowledgments:}

 The author thanks the Organizers of the conference for the warm hospitality.

%
%



\begin{thebibliography}{9}

\bibitem{Eto:2008yi}
  M.~Eto, T.~Fujimori, S.~B.~Gudnason,
K.~Konishi, M.~Nitta, K.~Ohashi and W.~Vinci,
  ``Constructing Non-Abelian Vortices with Arbitrary Gauge Groups,''
  arXiv:0802.1020 [hep-th].


\bibitem{Achucarro:1999it}
  A.~Achucarro and T.~Vachaspati,
  Phys.\ Rept.\  {\bf 327}, 347 (2000)
  [arXiv:hep-ph/9904229];
  R.~Jeannerot, J.~Rocher and M.~Sakellariadou,
  Phys.\ Rev.\ D {\bf 68}, 103514 (2003)
  [arXiv:hep-ph/0308134].


\bibitem{Vilenkin}
A. Vilenkin and E. P. S. Shellard, Cosmic Strings
and Other Topological Defects, Cambridge Univ. Press
(1994); M. B. Hindmarsh and T. W. B. Kibble, Rept. \
Prog. \  Phys. \ {\bf 58}, 477 (1995).

\bibitem{'tHooft:1981ht}
  G.~'t Hooft,
  Nucl.\ Phys.\  B {\bf 190}, 455 (1981);
  S.~Mandelstam,
  Phys.\ Lett.\  B {\bf 53}, 476 (1975).
\bibitem{Polyakov:1975yp}
  A.~M.~Polyakov and A.~A.~Belavin,
  JETP Lett.\  {\bf 22}, 245 (1975)
  [Pisma Zh.\ Eksp.\ Teor.\ Fiz.\  {\bf 22}, 503 (1975)].


\bibitem{Hanany:2003hp}
  A.~Hanany and D.~Tong,
  JHEP {\bf 0307}, 037 (2003)
  [arXiv:hep-th/0306150].

\bibitem{Auzzi:2003fs}
  R.~Auzzi, S.~Bolognesi, J.~Evslin, K.~Konishi and A.~Yung,
  Nucl.\ Phys.\  B {\bf 673}, 187 (2003)
  [arXiv:hep-th/0307287].


\bibitem{Abrikosov:1956sx}
A.~A.~Abrikosov,
Sov.\ Phys.\ JETP {\bf 5} (1957) 1174
[Zh.\ Eksp.\ Teor.\ Fiz.\ {\bf 32} (1957) 1442];
H.~B.~Nielsen and P.~Olesen,
Nucl.\ Phys.\ B {\bf 61} (1973) 45.


\bibitem{Eto:2005yh}
  M.~Eto, Y.~Isozumi, M.~Nitta, K.~Ohashi and N.~Sakai,
  Phys.\ Rev.\ Lett.\  {\bf 96}, 161601 (2006)
  [arXiv:hep-th/0511088];
  M.~Eto, K.~Konishi, G.~Marmorini, M.~Nitta, K.~Ohashi, W.~Vinci and N.~Yokoi,
  Phys.\ Rev.\  D {\bf 74}, 065021 (2006) 
  [arXiv:hep-th/0607070].
\bibitem{Eto:2006db}
  M.~Eto, 
  K.~Hashimoto, G.~Marmorini, M.~Nitta, K.~Ohashi and W.~Vinci,
  Phys.\ Rev.\ Lett.\  {\bf 98}, 091602 (2007) 
  [arXiv:hep-th/0609214];
  M.~Eto, 
  Y.~Isozumi, M.~Nitta, K.~Ohashi and N.~Sakai,
  J.\ Phys.\ A  {\bf 39}, R315 (2006)
  [arXiv:hep-th/0602170];
  M.~Eto, 
  T.~Fujimori, M.~Nitta, K.~Ohashi, K.~Ohta and N.~Sakai,
  Nucl.\ Phys.\  B {\bf 788}, 120 (2008) 
  [arXiv:hep-th/0703197].
 
\bibitem{DKO} D.~Dorigoni, K.~Konishi and K.~Ohashi, arXiv:0801.3284 [hep-th] (2008).
\bibitem{Ferretti:2007rp}
  L.~Ferretti, S.~B.~Gudnason and K.~Konishi,
  Nucl.\ Phys.\  B {\bf 789}, 84 (2008)
\bibitem{Goddard:1976qe}
  P.~Goddard, J.~Nuyts and D.~I.~Olive,
  Nucl.\ Phys.\  B {\bf 125}, 1 (1977).

\bibitem{Shifman:2004dr}
  M.~Shifman and A.~Yung,
  Phys.\ Rev.\  D {\bf 70}, 045004 (2004)
  [arXiv:hep-th/0403149];
  A.~Hanany and D.~Tong,
  JHEP {\bf 0404}, 066 (2004)
  [arXiv:hep-th/0403158].
  M.~Eto, 
  Y.~Isozumi, M.~Nitta, K.~Ohashi and N.~Sakai,
  Phys.\ Rev.\  D {\bf 72}, 025011 (2005)
  [arXiv:hep-th/0412048].


\bibitem{Tong:2005un}
  D.~Tong,
  arXiv:hep-th/0509216;
  M.~Shifman and A.~Yung,
  arXiv:hep-th/0703267.



\bibitem{Isozumi:2004vg}
  Y.~Isozumi, M.~Nitta, K.~Ohashi and N.~Sakai,
  Phys.\ Rev.\  D {\bf 71}, 065018 (2005)
  [arXiv:hep-th/0405129];
  Phys.\ Rev.\ Lett.\  {\bf 93}, 161601 (2004)
  [arXiv:hep-th/0404198];
  Phys.\ Rev.\  D {\bf 70}, 125014 (2004)
  [arXiv:hep-th/0405194].


\bibitem{Shifman:2006kd}
  M.~Shifman and A.~Yung,
  Phys.\ Rev.\  D {\bf 73}, 125012 (2006)
  [arXiv:hep-th/0603134];
M.~Eto, 
   J.~Evslin, K.~Konishi, G.~Marmorini, M.~Nitta, K.~Ohashi, W.~Vinci 
   and  N.~Yokoi,
  Phys.\ Rev.\  D {\bf 76}, 105002 (2007) 
  [arXiv:0704.2218 [hep-th]].

\bibitem{Eto:2006dx}
M.~Eto, L.~Ferretti, K.~Konishi, G.~Marmorini, M.~Nitta, K.~Ohashi, W.~Vinci 
   and  N.~Yokoi,
  Nucl.\ Phys.\  B {\bf 780}, 161 (2007)
  [arXiv:hep-th/0611313].


\bibitem{Eto:2008qw}
  M.~Eto, T.~Fujimori, S.~B.~Gudnason, M.~Nitta and K.~Ohashi,
  arXiv:0809.2014 [hep-th].


\bibitem{EFKGNOW}
  M.~Eto, T.~Fujimori, K.~Konishi, S.~B.~Gudnason, M.~Nitta and K.~Ohashi, W.~Vinci, work in preparation.



\bibitem{Luty:1995sd}
  M.~A.~Luty and W.~Taylor,
  Phys.\ Rev.\  D {\bf 53}, 3399 (1996).



\bibitem{Cvitanovic:1976am}
  P.~Cvitanovic,
  Phys.\ Rev.\  D {\bf 14}, 1536 (1976).



\bibitem{Hindmarsh:1991jq}
  M.~Hindmarsh,
  Phys.\ Rev.\ Lett.\  {\bf 68}, 1263 (1992).

\bibitem{Vachaspati:1991dz}
  T.~Vachaspati and A.~Achucarro,
  Phys.\ Rev.\  D {\bf 44}, 3067 (1991);
\bibitem{Preskill:1992bf}
  J.~Preskill,
  Phys.\ Rev.\  D {\bf 46}, 4218 (1992)
  [arXiv:hep-ph/9206216].

\end{thebibliography}
\end{document}